\documentclass[aps,prl,twocolumn,groupedaddress]{revtex4}
\usepackage{graphicx}

\newcommand{\omegaL}{\omega_\mathrm{L}}

\newcommand{\omegaS}{\omega_\mathrm{S}}

\newcommand{\ELO}{E_\mathrm{LO}}
\newcommand{\EO}{E}
\newcommand{\EL}{E_\mathrm{L}}

\begin{document}


\title{Wide-field Fourier transform spectral imaging}

\author{Michael Atlan}
\author{Michel Gross}
\affiliation{Laboratoire Kastler Brossel, \'Ecole Normale
Sup\'erieure, Universit\'e Pierre et Marie-Curie - Paris 6, Centre
National de la Recherche Scientifique, UMR 8552; 24 rue Lhomond,
75005 Paris, France}

\date{\today}

\begin{abstract}
We report experimental results of parallel measurement of spectral
components of the light. The temporal fluctuations of an optical
field mixed with a separate reference are recorded with a high
throughput complementary metal oxide semi-conductor camera (1
Megapixel at 2 kHz framerate). A numerical Fourier transform of the
time-domain recording enables wide-field coherent spectral imaging.
Qualitative comparisons with frequency-domain wide-field laser
Doppler imaging are provided.
\end{abstract}



\maketitle

Many coherent spectral detection schemes using a single detector (or
balanced detection) to detect temporal fluctuation spectra in an
optical mixing configuration rely on Fourier Transform spectroscopy
(FTS) for signal measurement \cite{Pike1970, Chung1997}. They
provide a high spectral resolution and shot-noise sensitivity. They
allow to shift away the $1/f$ noise of laser intensity fluctuations
since the measurement is done with GHz-bandwidth detectors. Most
imaging configurations require a spatial scanning of the beam, but
two approaches to parallel coherent spectral imaging with a
solid-state array detector were presented recently : full-field
laser Doppler imaging (LDI) \cite{Serov2002, SerovLasser2005,
Serov2005} and frequency-domain wide-field LDI (FDLDI)
\cite{AtlanGrossVitalis2006, AtlanGross2006, AtlanGross2007JOSAA}.
In the former approach, the temporal fluctuations of an optical
object field impinging on a complementary metal oxide semi-conductor
(CMOS) camera are recorded. Spectral imaging is done by calculating
the intensity-fluctuation spectrum by a Fourier transform (FT). One
major weakness of this approach lies in its inapplicability in
low-light conditions. The latter approach uses a spatiotemporal
heterodyne detection, which consists in recording an optical mix of
the object field with an angularly tilted and frequency-shifted
local oscillator (LO). It enables to measure spectral maps with a
high sensitivity but requires to acquire the spectral components
sequentially by sweeping the LO frequency. We present an alternative
approach, designed to combine the advantages of both methods. It
uses the properties of digital off-axis holography and FTS to enable
exploring of the temporal frequency spectrum of the object field.
Basically, the parallel spectral imaging instrument presented here
uses a CMOS camera to record the intensity fluctuations of an object
field mixed with a separate reference (LO); the field spectral
components are calculated by FTS.

\begin{figure}[]
\centering
\includegraphics[width=7.5cm]{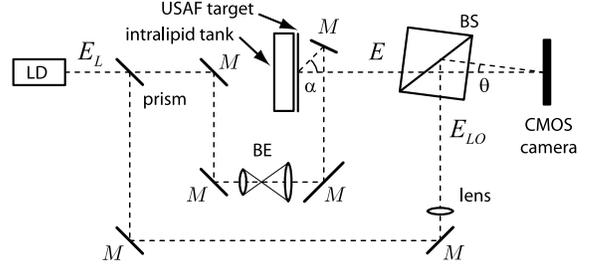}
\caption{Setup. LD : single mode laser diode. $M$ : mirror. BE :
beam expander. BS : beam splitter. $\EO$ : object field. $\ELO$ :
local oscillator field.}\label{fig_setup}
\end{figure}

The experimental setup is based on an optical interferometer
sketched in Fig.\ref{fig_setup}. A CW, 80 mW, $\lambda = $ 658 nm
diode (Mitsubishi ML120G21) provides the main laser beam (field
$\EL$, angular frequency $\omegaL$). A small part of this beam is
split by a prism to form a reference (LO) beam, while the remaining
part is expanded and illuminates an object in reflection with an
average incidence angle $\alpha \approx 45 ^\circ$. The object is
made of a USAF 1951 target set in front of a 4 mm-thick transparent
tank, filled with a non dilute intralipid (TM) 10\% emulsion. To
benefit from heterodyne gain, the field scattered by the object,
$\EO$, is mixed with the LO field $\ELO$ ( $ | \ELO |^2 / | \EO |^2
\sim 10^3)$, and is detected by a CMOS camera (LaVision
HighSpeedStar 4,  10 bit, 1024 $\times$ 1024 pixels at $\omegaS /(2
\pi)$ = 2.0 kHz frame rate, pixel area ${d_{pix}}^2$ with
$d_{pix}=17.5 ~\mu \rm m$, set at a distance $d$ = 50 cm from the
object. A 10 mm focal length lens is placed in the reference arm in
order to create an off-axis ($\theta \approx 1^{\circ}$ tilt angle)
virtual point source in the object plane. This configuration
constitutes a lensless Fourier holographic setup \cite{Stroke1965}.

In the detector plane, the  LO and object fields are:
\begin{eqnarray}\label{eq_E_LO_field}
\nonumber
 \ELO (t) = {\cal E}_{LO}  e^{ i \omegaL t } +
\textrm{c.c.}\\
 E (x,y,t) = {\cal E}(x,y,t) e^{ i \omegaL t } +
\textrm{c.c.}
\end{eqnarray}
where ${\rm c.c.}$ is the complex conjugate term. The LO beam is a
spherical wave propagating along $z$, and thus the LO field envelope
${\cal E}_{LO}$ does not depend on $x,y,t$. The object field
envelope ${\cal E}$, which contains information on the object shape,
and which may exhibit speckle, depends on position $x,y$. It also
depends on time $t$ because of dynamic scattering. The intensity $I$
recorded by the camera can be expressed as a function of the complex
fields :
\begin{eqnarray}\label{eq_i_defn}
   I(x,y,t) =\overline{ \left| \EO(x,y,t) + \ELO(t) \right|^2}\\
   \nonumber =|{\cal E}(x,y,t)|^2+|{\cal E}_{LO}|^2 \\
   \nonumber +{\cal E}(x,y,t){\cal E}^*_{LO}+ {\cal E}^*(x,y,t){\cal E}_{LO}
\end{eqnarray}
where $\overline{A}$ is the time average of $A$ over the optical
period.
\begin{figure}[h]
\centering
\includegraphics[width=7.5cm]{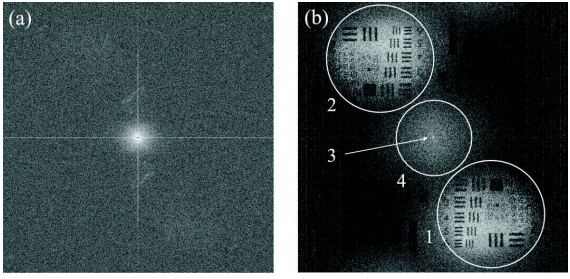}
\caption{Lensless off-axis holograms reconstructed in the target
plane. Images represent the field intensity, displayed in arbitrary
logarithmic scale. (a) image obtained from a single hologram. (b)
image obtained from the difference of two
holograms.}\label{fig_image_1ph_4ph}
\end{figure}
The camera records the interference pattern of the LO field ${\cal
E}_{LO}$ with the signal field $\cal E$, the recorded signal
$I(x,y,t)$ is the numerical hologram of the object that can be
used to reconstruct the object image \cite{SchnarsJuptner1994}.
Because of the lensless Fourier holographic configuration, the
reconstructed image field amplitude $\widetilde {\cal E}$ is
obtained from $I$ by a two-dimensional (2D) FT \cite{Wagner1999,
Schnars2002, Kreis2002}:
\begin{eqnarray}\label{eq_FFT2D}
    \widetilde {\cal E}(k_x,k_y)=\textrm{FT}^{2D}~I(x,y)
\end{eqnarray}
Fig.\ref{fig_image_1ph_4ph} shows intensity images of the USAF
target (i.e.$|~\widetilde{\cal E}(k_x,k_y)|^2$) displayed in
logarithmic scale (arbitrary units). Fig.\ref{fig_image_1ph_4ph}a is
obtained from a single frame $I(t_1)$ recorded at time $t_1$. The
USAF target is not visible because the noise is too large. To lower
the noise, we have recorded two frames $I(t_1)$ and $I(t_2)$ at
instants $t_1$ and $t_2$, and substracted them. By making the
difference of the two holograms, the noise components which do not
vary with time (like the LO beam noise and the CMOS dark signal
noise) cancel-out, whereas the Eq.\ref{eq_i_defn} holographic cross
terms (${\cal E}{\cal E}^*_{LO}$ and ${\cal E}^*{\cal E}_{LO}$) do
not vanish, because the signal field envelopes ${\cal E}(t_1)$ and
${\cal E}(t_2)$ are (at least partially) decorrelated in both
amplitude and phase from one frame to another as a consequence of
dynamic backscattering by the intralipid emulsion.
Fig.\ref{fig_image_1ph_4ph}b shows the reconstructed intensity image
($|\widetilde {\cal E}'|^2$) obtained from the difference of two
frames. One can notice that the last 4 terms of Eq.\ref{eq_i_defn}
are visible on Fig.\ref{fig_image_1ph_4ph}b. The true image (white
circle 1) corresponds to the cross term ${\cal E}{\cal E}^*_{LO}$,
while the twin image (white circle 2) corresponds to ${\cal
E}^*{\cal E}_{LO}$. Because of the lensless configuration, the true
and twin images are on focus in the same reconstruction plane (i.e.
the reciprocal plane of the detector). To prevent overlapping of the
true, twin and zero order images in the off-axis holographic
configuration, the true image size (circle 1 of diameter 0.75 cm 409
pixels) is $\sim 2.5\times$ smaller than the total 1024 pixels field
corresponding to $\lambda d / d_{pix}=1.87$ cm
\cite{SchnarsJuptner1994}. This means that, on average, one speckle
grain is about 2.5 pixels. The light collection efficiency is $2.5^2
\simeq 6 \times$ lower than with on-axis (or inline) holography or
with homodyne detection (for which one speckle = 1 pixel). $|{\cal
E}_{LO}|^2$ contributes to the zero order image
\cite{SchnarsJuptner1994, Cuche2000}. It yields the very bright
region in the center (null spatial frequency) of
Fig.\ref{fig_image_1ph_4ph}a and Fig.\ref{fig_image_1ph_4ph}b (arrow
3). Contrarily to $|{\cal E}_{LO}|^2$, the  $|{\cal E}|^2$ term is
not flat-field. It yields the broad spot in the center of
Fig.\ref{fig_image_1ph_4ph}b (circle 4). Because the brownian
spectrum is narrower than the Nyquist frequency of the time-domain
sampling, $\cal E$ and $I$ vary not too fast in time to be sampled
properly. It is then possible to record with the CMOS camera the
time evolution of intensity fluctuations in time. From a sequence of
CMOS images, one can thus extract the Fourier temporal frequency
components of the holographic signal, and reconstruct images from
these spectral components.
\begin{figure}[]
\centering
\includegraphics[width=7.5cm]{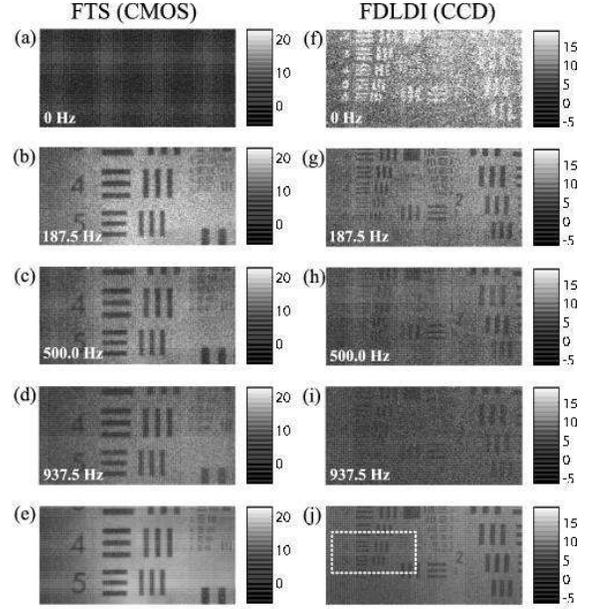}
\caption{$128 \times 256$ pixels images of temporal frequency
components of the object field, measured with the presented FTS
instrument using a CMOS detector: (a) to (d), and the wide field
FDLDI setup using a CCD detector: (f) to (i). (e) and (j): average
over frequencies. Color axis is in logarithmic arbitrary
units.}\label{fig_051013maps}
\end{figure}
We have recorded a data cube made of a sequence of $N$~=~2048 images
at a framerate $\omegaS / (2 \pi)$ = 2 kHz. A 3D numerical FT (2D
for space, 1D for time) was applied to this data to calculate
spectral component maps of the object field envelope in the target
plane :
\begin{eqnarray}\label{eq_fft3d}
\widetilde {\widetilde{\cal E}}(k_x,k_y,\omega)=
\textrm{FT}^{1D}~\widetilde{\cal E}(k_x,k_y,t)=
\textrm{FT}^{3D}~{I}(x,y,t)
\end{eqnarray}
The FT along the temporal dimension is used to calculate spectral
maps of the object field in quadrature (amplitude and phase), and
the FT in the spatial dimensions yields the field distribution in
the object plane (image). Since we have performed a discrete FT, the
2048 frequency points $\omega$ are linearly spaced between the
Nyquist frequencies $\pm 1.0 ~ \rm kHz$. The measurement time of the
$1024 \times 1024 \times 2048$ data cube is $\simeq 1$ s, and the
$\textrm{FT}^{\textrm{3D}}$ calculation time on a personal computer
is about 1 hour nowadays. Fig.\ref{fig_051013maps}(a) to
\ref{fig_051013maps}(d) show the images of the object field
intensity $|\widetilde{\widetilde{\cal E}}|^2$ in the target plane
for the frequency components $\omega/(2\pi) = 0$ (a), 187.5 (b),
500.0 (c) and 937.5 Hz (d) ($128 \times 256$ pixels crops of the
total hologram, displayed in logarithmic scale). For $\omega = 0$
(a), the LO beam noise is dominant and the target is not visible.
For $\omega \ne 0$, the USAF target is visible but the brightness
and SNR of the image will decrease with frequency ((b) to (d)).
Fig.\ref{fig_051013maps}(e) shows the image obtained by averaging
over all frequencies.
We have compared these results with wide-field FDLDI images
\cite{AtlanGrossVitalis2006, AtlanGross2006} obtained with a
charge-coupled device (CCD) camera (PCO Pixelfly: $1280\times 1024$
pixels, framerate: 8 Hz) with four-phase demodulation over 32 images
per spectral point, in the same experiment. Fig.\ref{fig_051013maps}
shows the $128 \times 256$ pixels FDLDI images at 0 (f), 187.5 (g),
500.0 (h) and 937.5 Hz (i), while image (j) corresponds to the
average over all frequencies. The USAF target is seen on all the
images. For $\omega = 0$,  the target appears as a contrast-reversed
image \cite{AtlanGrossVitalis2006}. Since the pixel size of the CCD
camera ($6.7 \times 6.7 ~ \mu \rm m $) is smaller than its CMOS
counterpart ($17.5 \mu {\rm m} \times 17.5 \mu {\rm m}$), the
extension of FDLDI image is larger (the acceptance angle of the
receiver is proportional to the inverse of the pixel size). The
white dashed rectangle of Fig.\ref{fig_051013maps}j corresponds to
the CMOS-imager field of view. Although the number of recorded
spectral points was kept low for the FDLDI measurement compared to
the FTS scheme (64 vs. 2048), the total measurement time was much
greater (256 seconds vs. 1 second). This difference is due to the
throughput discrepancy between the CCD (1.3 Mpixel @ 8 Hz) and the
CMOS (1.0 Mpixel @ 2 kHz) receivers.
\begin{figure}[]
\centering
\includegraphics[width = 6.5 cm]{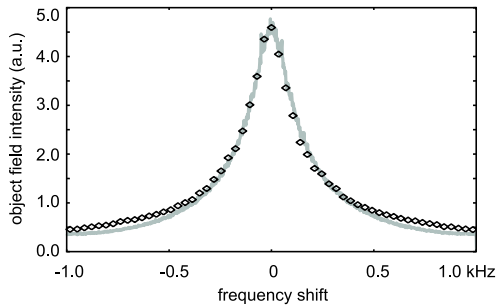}
\caption{Spectrum of the field dynamically backscattered by a non
dilute suspension of intralipid 10\% obtained by averaging the
object field intensity over 50 $\times$ 50 pixels. FTS (solid gray
curve) and FDLDI (points) spectra. Horizontal axis is the frequency
$\omega/(2 \pi)$ in kHz, vertical axis is signal in linear arbitrary
units.}\label{fig_051013spectra_comp_5x5_50x50}
\end{figure}

We have computed the frequency spectrum of the light diffused by the
intralipid emulsion with FTS. This spectrum is obtained by averaging
the object field intensity over a $50 \times 50 $ pixels region of
the reconstructed image. The lineshape is plotted on
Fig.\ref{fig_051013spectra_comp_5x5_50x50} as a solid gray curve. We
have compared its shape with the one obtained with the FDLDI
technique (Fig.\ref{fig_051013spectra_comp_5x5_50x50} points). The
agreement is good except in the tails of the spectrum. The FTS
frequency response is imperfectly flat, because of the CCD finite
exposure time ($1/\omega_S$) that yields signal low pass filtering
\cite{PicartLeval2003, AtlanGross2007}. Moreover, because the signal
temporal evolution is sampled at 2 kHz, temporal sampling aliases
and spectrum overlap are expected around the Nyquist frequencies
$\pm 1$ kHz.

In this Letter, we have shown that the spatiotemporal heterodyne
detection recently introduced \cite{AtlanGross2006,
AtlanGrossVitalis2006, AtlanGross2007JOSAA} can be adapted to a
wide-field Fourier transform spectral imaging scheme with a high
throughput array detector. By using an off-axis optical mixing
configuration, the object-LO fields cross terms are shifted away
from center of the detector reciprocal plane (k-space), contrarily
to the object and LO self-beating contributions, which remain
unshifted. It is then possible to reject the local oscillator and
the object field self-beating contributions accounting for noise.
The heterodyne gain provided by optical amplification of the object
field by the LO field is essential for a high frame rate camera
measurement in low-light conditions, since the object field
intensity decreases with the camera exposure time. The ability to
filter-off the LO beam noise, yields an optimal sensitivity of 1
photoelectron of noise per pixel. This limit has been reached with
4-phase detection \cite{GrossAtlan2007}, which consists of a
discrete Fourier transform on 4 data points to calculate a single
frequency component of the object field. Here, the expected noise
limit is the same for each frequency component of the object field
obtained by discrete Fourier transform. This method might find
applications in dynamic light scattering analysis of colloidal
suspensions and microfluidic systems.



\begin{thebibliography}{17}
\expandafter\ifx\csname
natexlab\endcsname\relax\def\natexlab#1{#1}\fi
\expandafter\ifx\csname bibnamefont\endcsname\relax
  \def\bibnamefont#1{#1}\fi
\expandafter\ifx\csname bibfnamefont\endcsname\relax
  \def\bibfnamefont#1{#1}\fi
\expandafter\ifx\csname citenamefont\endcsname\relax
  \def\citenamefont#1{#1}\fi
\expandafter\ifx\csname url\endcsname\relax
  \def\url#1{\texttt{#1}}\fi
\expandafter\ifx\csname urlprefix\endcsname\relax\def\urlprefix{URL
}\fi \providecommand{\bibinfo}[2]{#2}
\providecommand{\eprint}[2][]{\url{#2}}

\bibitem[{\citenamefont{Pike}(1970)}]{Pike1970}
\bibinfo{author}{\bibfnamefont{E.}~\bibnamefont{Pike}},
  \bibinfo{journal}{Review of Physics in Technology}
  \textbf{\bibinfo{volume}{1}}, \bibinfo{pages}{180} (\bibinfo{year}{1970}).

\bibitem[{\citenamefont{Chung et~al.}(1997)\citenamefont{Chung, Lee, and
  Mazur}}]{Chung1997}
\bibinfo{author}{\bibfnamefont{D.}~\bibnamefont{Chung}},
  \bibinfo{author}{\bibfnamefont{K.}~\bibnamefont{Lee}}, \bibnamefont{and}
  \bibinfo{author}{\bibfnamefont{E.}~\bibnamefont{Mazur}},
  \bibinfo{journal}{Applied physics. B, Lasers and optics}
  \textbf{\bibinfo{volume}{64}}, \bibinfo{pages}{1} (\bibinfo{year}{1997}).

\bibitem[{\citenamefont{Serov et~al.}(2002)\citenamefont{Serov, Steenbergen,
  and de~Mul}}]{Serov2002}
\bibinfo{author}{\bibfnamefont{A.}~\bibnamefont{Serov}},
  \bibinfo{author}{\bibfnamefont{W.}~\bibnamefont{Steenbergen}},
  \bibnamefont{and} \bibinfo{author}{\bibfnamefont{F.}~\bibnamefont{de~Mul}},
  \bibinfo{journal}{Optics Letters} \textbf{\bibinfo{volume}{27}},
  \bibinfo{pages}{300} (\bibinfo{year}{2002}).

\bibitem[{\citenamefont{Serov and Lasser}(2005)}]{SerovLasser2005}
\bibinfo{author}{\bibfnamefont{A.}~\bibnamefont{Serov}} \bibnamefont{and}
  \bibinfo{author}{\bibfnamefont{T.}~\bibnamefont{Lasser}},
  \bibinfo{journal}{Opt. Express} \textbf{\bibinfo{volume}{13}},
  \bibinfo{pages}{6416} (\bibinfo{year}{2005}).

\bibitem[{\citenamefont{Serov et~al.}(2005)\citenamefont{Serov, Steinacher, and
  Lasser}}]{Serov2005}
\bibinfo{author}{\bibfnamefont{A.}~\bibnamefont{Serov}},
  \bibinfo{author}{\bibfnamefont{B.}~\bibnamefont{Steinacher}},
  \bibnamefont{and} \bibinfo{author}{\bibfnamefont{T.}~\bibnamefont{Lasser}},
  \bibinfo{journal}{Opt. Ex.} \textbf{\bibinfo{volume}{13}},
  \bibinfo{pages}{3681} (\bibinfo{year}{2005}).

\bibitem[{\citenamefont{Atlan et~al.}(2006)\citenamefont{Atlan, Gross, Vitalis,
  Rancillac, Forget, and Dunn}}]{AtlanGrossVitalis2006}
\bibinfo{author}{\bibfnamefont{M.}~\bibnamefont{Atlan}},
  \bibinfo{author}{\bibfnamefont{M.}~\bibnamefont{Gross}},
  \bibinfo{author}{\bibfnamefont{T.}~\bibnamefont{Vitalis}},
  \bibinfo{author}{\bibfnamefont{A.}~\bibnamefont{Rancillac}},
  \bibinfo{author}{\bibfnamefont{B.~C.} \bibnamefont{Forget}},
  \bibnamefont{and} \bibinfo{author}{\bibfnamefont{A.~K.} \bibnamefont{Dunn}},
  \bibinfo{journal}{Optics Letters} \textbf{\bibinfo{volume}{31}}
  (\bibinfo{year}{2006}).

\bibitem[{\citenamefont{Atlan and Gross}(2006)}]{AtlanGross2006}
\bibinfo{author}{\bibfnamefont{M.}~\bibnamefont{Atlan}} \bibnamefont{and}
  \bibinfo{author}{\bibfnamefont{M.}~\bibnamefont{Gross}},
  \bibinfo{journal}{Review of Scientific Instruments}
  \textbf{\bibinfo{volume}{77}}, \bibinfo{pages}{1161031}
  (\bibinfo{year}{2006}).

\bibitem[{\citenamefont{Atlan and Gross}(2007)}]{AtlanGross2007JOSAA}
\bibinfo{author}{\bibfnamefont{M.}~\bibnamefont{Atlan}} \bibnamefont{and}
  \bibinfo{author}{\bibfnamefont{M.}~\bibnamefont{Gross}},
  \bibinfo{journal}{Journal of the Optical Society of America A}
  \textbf{\bibinfo{volume}{24}}, \bibinfo{pages}{2701} (\bibinfo{year}{2007}).

\bibitem[{\citenamefont{Stroke}(1965)}]{Stroke1965}
\bibinfo{author}{\bibfnamefont{G.~W.} \bibnamefont{Stroke}},
  \bibinfo{journal}{Applied Physics Letters} \textbf{\bibinfo{volume}{6}},
  \bibinfo{pages}{201} (\bibinfo{year}{1965}).

\bibitem[{\citenamefont{Schnars and Juptner}(1994)}]{SchnarsJuptner1994}
\bibinfo{author}{\bibfnamefont{U.}~\bibnamefont{Schnars}} \bibnamefont{and}
  \bibinfo{author}{\bibfnamefont{W.}~\bibnamefont{Juptner}},
  \bibinfo{journal}{Appl. Opt.} \textbf{\bibinfo{volume}{33}},
  \bibinfo{pages}{179} (\bibinfo{year}{1994}).

\bibitem[{\citenamefont{Wagner et~al.}(1999)\citenamefont{Wagner, Seebacher,
  Osten, and Juptner}}]{Wagner1999}
\bibinfo{author}{\bibfnamefont{C.}~\bibnamefont{Wagner}},
  \bibinfo{author}{\bibfnamefont{S.}~\bibnamefont{Seebacher}},
  \bibinfo{author}{\bibfnamefont{W.}~\bibnamefont{Osten}}, \bibnamefont{and}
  \bibinfo{author}{\bibfnamefont{W.}~\bibnamefont{Juptner}},
  \bibinfo{journal}{Applied Optics} \textbf{\bibinfo{volume}{38}},
  \bibinfo{pages}{4812} (\bibinfo{year}{1999}).

\bibitem[{\citenamefont{Schnars and Juptner}(2002)}]{Schnars2002}
\bibinfo{author}{\bibfnamefont{U.}~\bibnamefont{Schnars}} \bibnamefont{and}
  \bibinfo{author}{\bibfnamefont{W.~P.~O.} \bibnamefont{Juptner}},
  \bibinfo{journal}{Meas. Sci. Technol.} \textbf{\bibinfo{volume}{13}},
  \bibinfo{pages}{R85} (\bibinfo{year}{2002}).

\bibitem[{\citenamefont{Kreis}(2002)}]{Kreis2002}
\bibinfo{author}{\bibfnamefont{T.~M.} \bibnamefont{Kreis}},
  \bibinfo{journal}{Optical Engineering} \textbf{\bibinfo{volume}{41}},
  \bibinfo{pages}{771} (\bibinfo{year}{2002}).

\bibitem[{\citenamefont{Cuche et~al.}(2000)\citenamefont{Cuche, Marquet, and
  Depeursinge}}]{Cuche2000}
\bibinfo{author}{\bibfnamefont{E.}~\bibnamefont{Cuche}},
  \bibinfo{author}{\bibfnamefont{P.}~\bibnamefont{Marquet}}, \bibnamefont{and}
  \bibinfo{author}{\bibfnamefont{C.}~\bibnamefont{Depeursinge}},
  \bibinfo{journal}{Applied Optics} \textbf{\bibinfo{volume}{39}},
  \bibinfo{pages}{4070} (\bibinfo{year}{2000}).

\bibitem[{\citenamefont{Picart et~al.}(2003)\citenamefont{Picart, Leval,
  Mounier, and Gougeon}}]{PicartLeval2003}
\bibinfo{author}{\bibfnamefont{P.}~\bibnamefont{Picart}},
  \bibinfo{author}{\bibfnamefont{J.}~\bibnamefont{Leval}},
  \bibinfo{author}{\bibfnamefont{D.}~\bibnamefont{Mounier}}, \bibnamefont{and}
  \bibinfo{author}{\bibfnamefont{S.}~\bibnamefont{Gougeon}},
  \bibinfo{journal}{Opt. Lett.} \textbf{\bibinfo{volume}{28}},
  \bibinfo{pages}{1900} (\bibinfo{year}{2003}).

\bibitem[{\citenamefont{Atlan et~al.}(2007)\citenamefont{Atlan, Gross, and
  Absil}}]{AtlanGross2007}
\bibinfo{author}{\bibfnamefont{M.}~\bibnamefont{Atlan}},
  \bibinfo{author}{\bibfnamefont{M.}~\bibnamefont{Gross}}, \bibnamefont{and}
  \bibinfo{author}{\bibfnamefont{E.}~\bibnamefont{Absil}},
  \bibinfo{journal}{Optics Letters} \textbf{\bibinfo{volume}{32}},
  \bibinfo{pages}{1456} (\bibinfo{year}{2007}).

\bibitem[{\citenamefont{Gross and Atlan}(2007)}]{GrossAtlan2007}
\bibinfo{author}{\bibfnamefont{M.}~\bibnamefont{Gross}} \bibnamefont{and}
  \bibinfo{author}{\bibfnamefont{M.}~\bibnamefont{Atlan}},
  \bibinfo{journal}{Optics Letters} \textbf{\bibinfo{volume}{32}},
  \bibinfo{pages}{909} (\bibinfo{year}{2007}).

\end{thebibliography}

\newpage

\end{document}